# Indexing the Sphere with the Hierarchical Triangular Mesh

Alexander S. Szalay[1],
Jim Gray[2],
George Fekete[1]
Peter Z. Kunszt[3],
Peter Kukol[2]
Ani Thakar[1],

1. The Johns Hopkins University
2. Microsoft Research
3. CERN, Geneva

August 2005

Technical Report

MSR-TR-2005-123

# Indexing the Sphere with the Hierarchical Triangular Mesh

Alexander S. Szalay[1], Jim Gray[2], George Fekete[1] Peter Z. Kunszt[3],
Peter Kukol[2], and Ani Thakar[1]

1. Dept of Physics and Astronomy, The Johns Hopkins University, Baltimore
2. Microsoft Research Bay Area Research Center, San Francisco
3. CERN, Geneva

**Abstract:** We describe a method to subdivide the surface of a sphere into spherical triangles of similar, but not identical, shapes and sizes. The Hierarchical Triangular Mesh (HTM) is a quad-tree that is particularly good at supporting searches at different resolutions, from arc seconds to hemispheres. The subdivision scheme is universal, providing the basis for addressing and for fast lookups. The HTM provides the basis for an efficient geospatial indexing scheme in relational databases where the data have an inherent location on either the celestial sphere or the Earth. The HTM index is superior to cartographical methods using coordinates with singularities at the poles. We also describe a way to specify surface regions that efficiently represent spherical query areas. This article presents the algorithms used to identify the HTM triangles covering such regions.

## 1. Introduction

Many science and commercial applications must catalog and search objects in three-dimensional space. In Earth and Space Science, the coordinate system is usually spherical, so the object's position is given by its location on the surface of the unit sphere and its distance from the origin. Queries on catalogs of such objects often involve constraints on these coordinates, often in terms of complex regions that imply complicated spherical trigonometry.

There is a great interest in a universal, computer-friendly index on the sphere, especially in astronomy, where the ancient index of stellar constellations is still in common use, and in earth sciences, where people use maps having complicated spherical projections. The spatial index presented here transforms regions of the sphere to unique identifiers (IDs). These IDs can be used both as an identifier for an area and as an indexing strategy. The transformation uses only elementary spherical geometry to identify a certain area. It provides universality, which is essential for cross-referencing across different datasets. The comparisons are especially well-suited for computers because they replace transcendental functions with a few multiplications and additions.

The technique to subdivide the sphere into spherical triangles presented here is recursive. At each level of recursion, the triangle areas are roughly the same (within a factor of 2), which is a major advantage over the usual spherical coordinate system projections with singularities at the poles. Also, in areas with high data density, the recursion process can be applied to a higher level than in areas where data points are rare. This enables uneven data distributions to be organized into equal-sized bins.

A similar scheme for subdividing the sphere was advocated by Barrett [1]. The idea of the current implementation of the HTM was described in Kunszt et al [2]. Short, Fekete et al [3,4,5] used an icosahedron-based mesh for Earth sciences applications. Quad trees and geometrical indexing are discussed in the books of Samet in detail [6,7]. Lee and Samet [8] use an identical triangulation, but a different numbering scheme. Goodchild [9,10] and Song et al [11] created a similar triangulation of the sphere, called the *Discrete Global Grid*, with precisely equal areas, using small circles for the hierarchical boundaries. Gray et al [12] described linking the HTM to a relational database system.

# 2. The Hierarchical Triangular Mesh

## 2.1. *Using Cartesian Coordinates*

Performing computations on the surface of the sphere generally involves complicated transcendental functions with singularities at the spherical coordinate system poles. In these coordinate systems, it is typically computationally expensive to evaluate even simple tests like point-inside-circle or circle-overlap-circle. The HTM approach uses a three-dimensional vector representation to circumvent these problems. By going from the two-dimensional spherical surface to three-dimensional volumes containing the surface area, point-in-polygon computations avoid transcendental computations and need just a few multiplications and additions, and a compare.

## 2.2. *Starting the Subdivision*

The hierarchical subdivision of the sphere starts with eight spherical triangles of equal size – the octahedron projected onto the sphere as illustrated by Figure 1. A spherical triangle is given by three points on the unit sphere connected by great circle segments. The octahedron has six vertices, given by the intersection points of the *x,y,z* axes with the unit sphere, which we enumerate $\mathbf{v}_0$ through $\mathbf{v}_5$:

$$\begin{aligned}
(0, 0, 1) &:= \mathbf{v}_0 \\
(1, 0, 0) &:= \mathbf{v}_1 \\
(0, 1, 0) &:= \mathbf{v}_2 \\
(-1, 0, 0) &:= \mathbf{v}_3 \\
(0, -1, 0) &:= \mathbf{v}_4 \\
(0, 0, -1) &:= \mathbf{v}_5
\end{aligned} \qquad (2.1)$$

The first eight nodes of the HTM index are defined as these eight spherical triangles named by using S for south and N for north, and then numbering the faces clockwise:

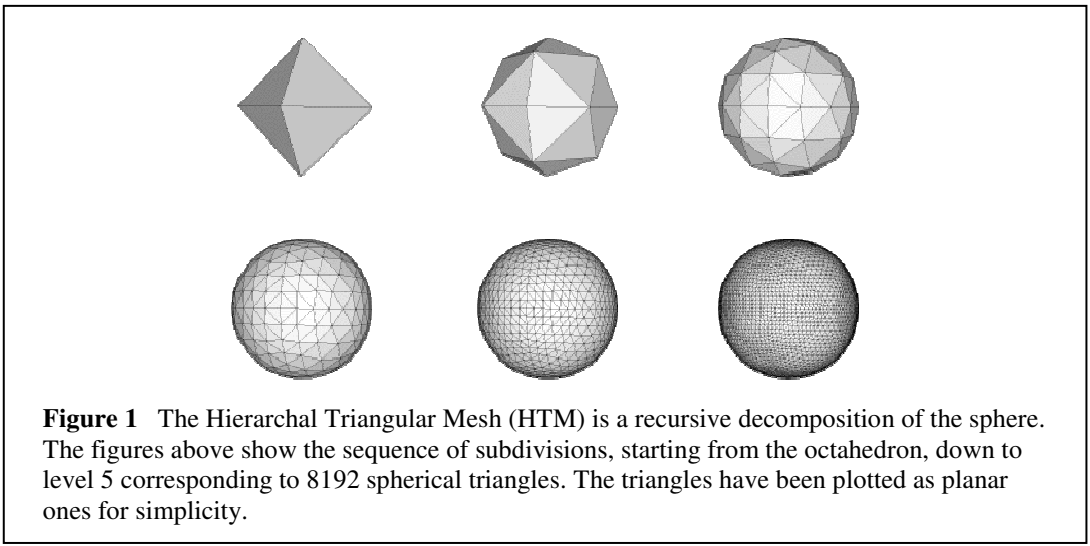

**Figure 1** The Hierarchal Triangular Mesh (HTM) is a recursive decomposition of the sphere. The figures above show the sequence of subdivisions, starting from the octahedron, down to level 5 corresponding to 8192 spherical triangles. The triangles have been plotted as planar ones for simplicity.

$$(\mathbf{v}_1, \mathbf{v}_5, \mathbf{v}_2) := S0, \quad (\mathbf{v}_1, \mathbf{v}_o, \mathbf{v}_4) := N0$$
$$(\mathbf{v}_2, \mathbf{v}_5, \mathbf{v}_3) := S1, \quad (\mathbf{v}_4, \mathbf{v}_o, \mathbf{v}_3) := N1$$
$$(\mathbf{v}_3, \mathbf{v}_5, \mathbf{v}_4) := S2, \quad (\mathbf{v}_3, \mathbf{v}_o, \mathbf{v}_2) := N2 \quad (2.2)$$
$$(\mathbf{v}_4, \mathbf{v}_5, \mathbf{v}_1) := S3, \quad (\mathbf{v}_2, \mathbf{v}_o, \mathbf{v}_1) := N3$$

The triangles are all given with the vertices traversed counterclockwise.

## 2.3. Recursive Subdivision of the Sphere

To get the next level of the recursion, for a particular triangle, we first rename the vertices to $(\mathbf{v}_0,\mathbf{v}_1,\mathbf{v}_2)$ by keeping their counterclockwise order, e.g. $S0: (\mathbf{v}_1,\mathbf{v}_5,\mathbf{v}_2) \rightarrow (\mathbf{v}_0,\mathbf{v}_1,\mathbf{v}_2)$. The triangle is subdivided into four smaller ones by $(\mathbf{w}_0,\mathbf{w}_1,\mathbf{w}_2)$, the midpoints of the edges:

$$\mathbf{w}_0 = \frac{\mathbf{v}_1 + \mathbf{v}_2}{|\mathbf{v}_1 + \mathbf{v}_2|}$$

$$\mathbf{w}_1 = \frac{\mathbf{v}_0 + \mathbf{v}_2}{|\mathbf{v}_0 + \mathbf{v}_2|} \quad (2.3)$$

$$\mathbf{w}_2 = \frac{\mathbf{v}_0 + \mathbf{v}_1}{|\mathbf{v}_0 + \mathbf{v}_1|}.$$

The four new triangles are given by

$$\text{Triangle } 0 := (\mathbf{v}_0, \mathbf{w}_2, \mathbf{w}_1),$$
$$\text{Triangle } 1 := (\mathbf{v}_1, \mathbf{w}_0, \mathbf{w}_2),$$
$$\text{Triangle } 2 := (\mathbf{v}_2, \mathbf{w}_1, \mathbf{w}_0), \quad (2.4)$$
$$\text{Triangle } 3 := (\mathbf{w}_0, \mathbf{w}_1, \mathbf{w}_2)$$

The node names of the new triangles are the name of the original triangle with the triangle number appended. If the original node name was *N2*, the new node names are *N20…N23* (Figure 2). To repeat the recursion, we rename the vertices of the four new triangles in (2.4) to ($v_0$,$v_1$,$v_2$), always keeping their order counter-clockwise. Note that this is different from the notation in [5], where the winding order is alternating.

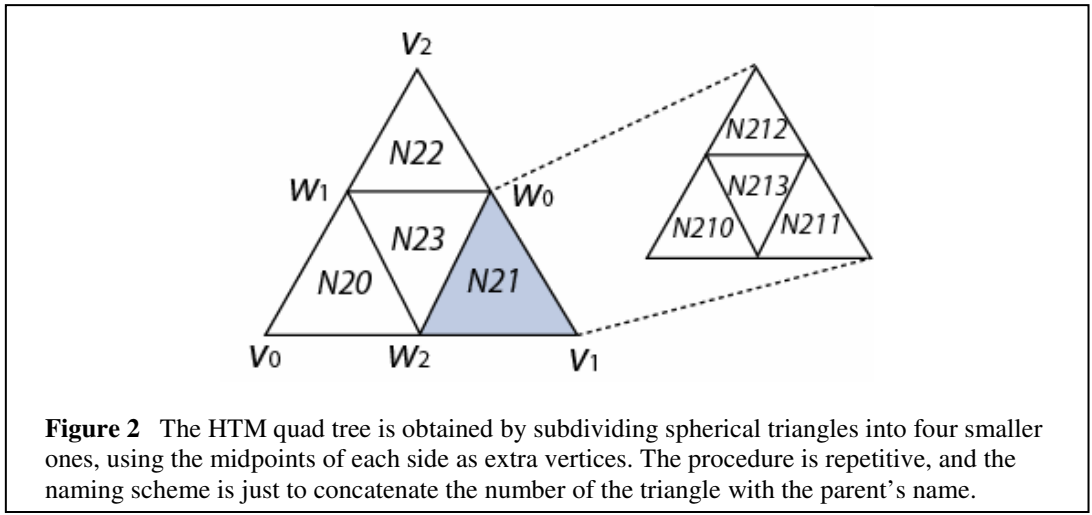

**Figure 2** The HTM quad tree is obtained by subdividing spherical triangles into four smaller ones, using the midpoints of each side as extra vertices. The procedure is repetitive, and the naming scheme is just to concatenate the number of the triangle with the parent's name.

The recursion can be repeated to any desired depth. The number of triangle nodes $N$ at a given depth $d>0$ is given by:

$$N(d) = 8 \times 4^{d-1}, \qquad (2.5)$$

where depth 0 corresponds to the whole sphere. The recursive process of equations (2.3, 2.4) define what we call the Hierarchical Triangular Mesh (HTM). The HTM is well suited to build a spatial index of 3D data that has an inherent spherical distribution by assigning the HtmID of a certain depth to every point in a given catalog.

## 2.4. Naming and Location

Each HTM triangle (at any depth) is referred to as a *trixel*. The name of a trixel (e.g. N204130012) uniquely defines its depth (number of digits in the name) and its location on the sky (name sequence). We can get an idea of the location of the trixel just by looking at its name. First of all, we immediately know which quadrant we are in (the first two characters), as given in Eq. (2). The next number gives its location in that quadrant – close to vertex number (0,1,2) or in the center (3). The numbers forming the name can be used to compute the exact location of the three vertex vectors of the given trixel.

The name of the HTM nodes can be easily encoded into an integer ID by assigning two bits to each level and encoding $N$ as 11 and $S$ as 10. For example, N01 encodes as binary 110001 or hexadecimal 0x31. A 64-bit integer can hold an HtmID up to depth 31. However, standard double precision transcendental functions break down at depth 26 where the trixel sides reach dimensions below $10^{-15}$ (depth 26 is about 10 milli-arcseconds for astronomers or 30 centimeters on the earth's surface.) The leading one bit signifies the depth of the tree. The southernmost face 0 node on the tree at any level has the binary form 10..00, thus the position of this leading bit is the only non-zero value and *depth(HtmID) = floor(log$_4$(HtmID))*.

Given a dataset of points, we can fix a depth $d$ and assign to each data point an HtmID. The HtmID to depth $d$ can be used as a key index in the given catalog. For example, the Sloan Digital Sky Survey and several other surveys use 21-deep HtmIDs to name distinct objects.

## 2.5. Properties and Statistics of the HTM

The number of nodes per recursion depth $d>0$ is $8 \times 4^{d-1}$ (see Eq. 2.5). Trixel areas at a given depth $d>0$ are scattered around the mean area of $\pi/(2 \times 4^{d-1})$. Figure 3 shows that at depth 7, about three-quarters of the trixels are slightly smaller than average size, and the remaining quarter are larger. This scatter is due to the difference in shapes—the three corner trixels are smaller than the center one. The ratio of the maximum over the minimum areas at higher depths is about 2. The smallest triangles are the ones adjacent to the vertices of the initial octahedron, while the largest ones are at the nested center triangles, with names Nx3333... or Sx3333… for x∈{0,1,2,3}. Beyond depth 7, the curvature becomes irrelevant, the distribution remains self-similar, and the scatter (variance) in the areas around the mean is 24 percent.

The shape difference is kept within limits; the maximal inner angle of a triangle never exceeds $\pi/2$ and is never less than $\pi/4$. The edges (arcs) forming the triangles are also within a fixed minimal and maximal value per level. The smallest side length encountered is always $\pi/2^n$– this is the subdivision of the side that lies on the edge of the original octahedron. All the other sides are longer. The maximal side length is always at the center of the level-0 octahedron triangle. The scatter (variance) in the arc lengths for the depth 7 triangles is 15 percent, and the ratio of the average arc length to the canonical size $\pi/2^n$ is about 1.23.

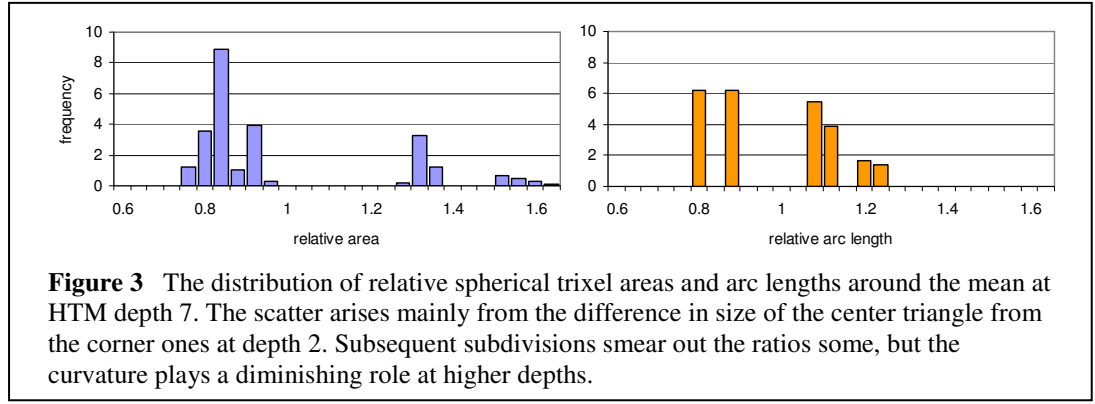

**Figure 3** The distribution of relative spherical trixel areas and arc lengths around the mean at HTM depth 7. The scatter arises mainly from the difference in size of the center triangle from the corner ones at depth 2. Subsequent subdivisions smear out the ratios some, but the curvature plays a diminishing role at higher depths.

# 3. Defining the Geometry

The HTM index must support intersections with arbitrary spherical regions. Given a region, we want to get all HTM trixels of a given depth that cover it. We call this set of trixels the region's *HTM cover*. This section presents the geometry primitives that define spherical areas. These primitives can easily be intersected with a trixel to categorize it as inside, outside, or partially overlapping the region. When the algorithms to represent and simplify a region are in place, we tackle the HTM cover computation.

## 3.1. Halfspaces

The *halfspace* is our basic geometry primitive. Each halfspace defines a cap, $h$, on the unit sphere that is inside the halfspace and so is sliced off by the plane that bounds the halfspace. Any halfspace can be characterized by the plane that bounds it and a direction from that plane. For our purposes, it is convenient to define the halfspace by (i) the vector, $\vec{v}$, from the origin pointing into the halfspace and normal to the halfspace's bounding plane, and (ii) the distance $d$ to the plane from the origin along the vector $\vec{v}$:

$$c := \{\vec{v}; d\} \qquad |\vec{v}| = 1 \; ; \; -1 \leq d \leq 1 \qquad (3.1)$$

Negative halfspaces, $d<0$, cover more than half of the spherical surface. They are holes on the sphere. The sign of the halfspace is defined to be the sign of $d$:

$$\text{sign}(c) := \text{sign}(d) \qquad (3.2)$$

We also define the arcangle of a halfspace:

$$\varphi_c = \arccos(d) \qquad (3.3)$$

*Examples*:

$\{\mathbf{v},d\}=(0,0,1,0.5)$ defines the cap north of 30°N latitude with angle 30°,
$\{\mathbf{v},d\}=(0,0,-1,-0.5)$ defines the area below 30°N latitude and has a 150° angle.

## 3.2. Convexes

We define a *convex*, $c$, as the intersection of halfspaces, $\{h_i\}$, i.e., the intersection of caps on the sphere:

$$c := h_1 \ \& \ h_2 \ \& \ \ldots \& \ h_n \quad n \in N^+ \quad (3.4)$$

The intersection volume is a (possibly open) convex polyhedron in 3D, but its intersection with the unit sphere is not necessarily a convex area or even a connected area. It is not even true that convexes are always a contiguous area on the sphere. Consider, for example, a cube in Figure 4 centered around the origin whose corner points are barely outside the unit sphere (e.g., each of the six planar halfspaces has offsets -0.6). The corresponding convex on the sphere is the eight triangular areas in the vicinity of each of the vertices of the cube. But, according to our definition, this is still a convex. These disjointed areas within a convex are called *patches*. Each patch is a connected region on the sphere.

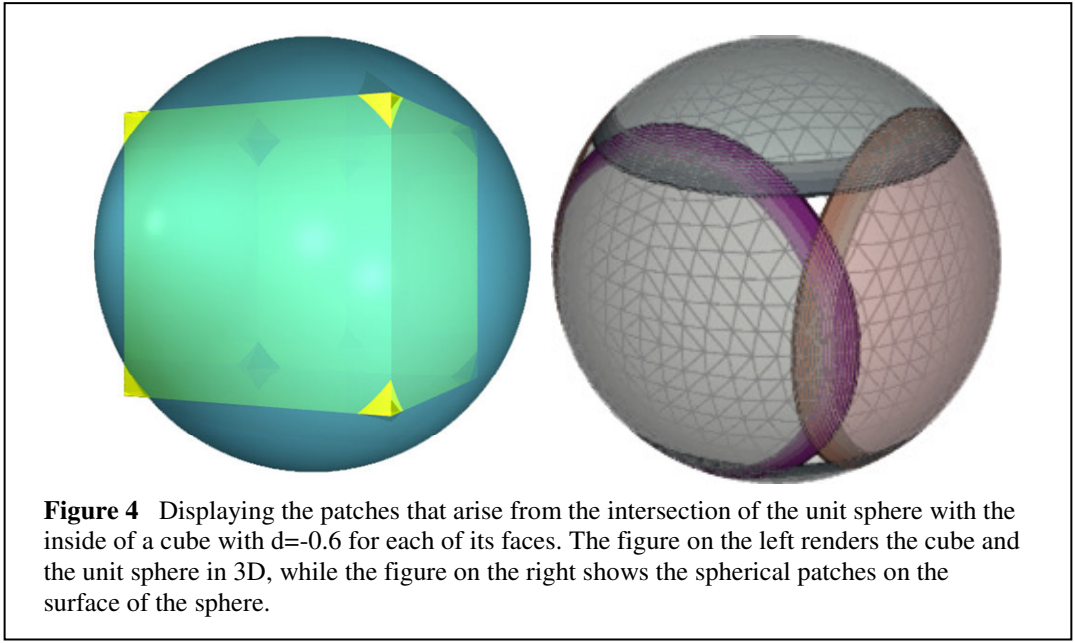

**Figure 4** Displaying the patches that arise from the intersection of the unit sphere with the inside of a cube with d=-0.6 for each of its faces. The figure on the left renders the cube and the unit sphere in 3D, while the figure on the right shows the spherical patches on the surface of the sphere.

Note that all these atypical convexes involve at least one negative halfspace. Convexes composed entirely of positive and zero halfspaces always define a convex area on the sphere that is a single patch. These non-negative convexes are easier to handle, so it makes sense to define a sign for the convexes, too. This sign will be important when we calculate the intersections with the HTM.

$$\text{sign}(x) := \begin{cases} -1, & \text{sign}(c_i) \text{ all -1 or 0} \\ 0, & \text{sign}(c_i) \text{ all 0} \\ +1, & \text{sign}(c_i) \text{ all +1 or 0} \\ mixed, & \text{at least one +1 and one -1} \end{cases} \quad (3.5)$$

## 3.3. Regions

We define a region, $r$, to be union of convexes $\{c_i\}$:

$$r := c_1 \ | \ c_2 \ | \ \ldots \ | \ c_n \quad n \in N^+ \quad (3.6)$$

One can think of regions as the disjunctive normal form of halfspaces—the union of a set of conjunctions.

Regions can represent very complicated spherical areas—anything that can be built up as a union of convex areas. If we devise a generic method to intersect regions with the HTM trixels, we have a powerful indexing tool at hand.

## 3.4. Trivial Simplifications

Several kinds of halfspaces can be trivially eliminated. These are performed immediately after assembling the halfspace array.

(i) *Duplicates*
If the same halfspace appears more than once, only one copy is kept.

(ii) *Complements*
If a halfspace and its exact complement appear, the convex is NULL.

(iii) *Whole Sphere*
A halfspace that contains the whole sphere, i.e., $d \leq -1$ can be dropped unless it is the only halfspace of the convex.

(iv) *Null*
If $d > 1$ for some halfspace $\{n,d\}$, the convex is NULL.

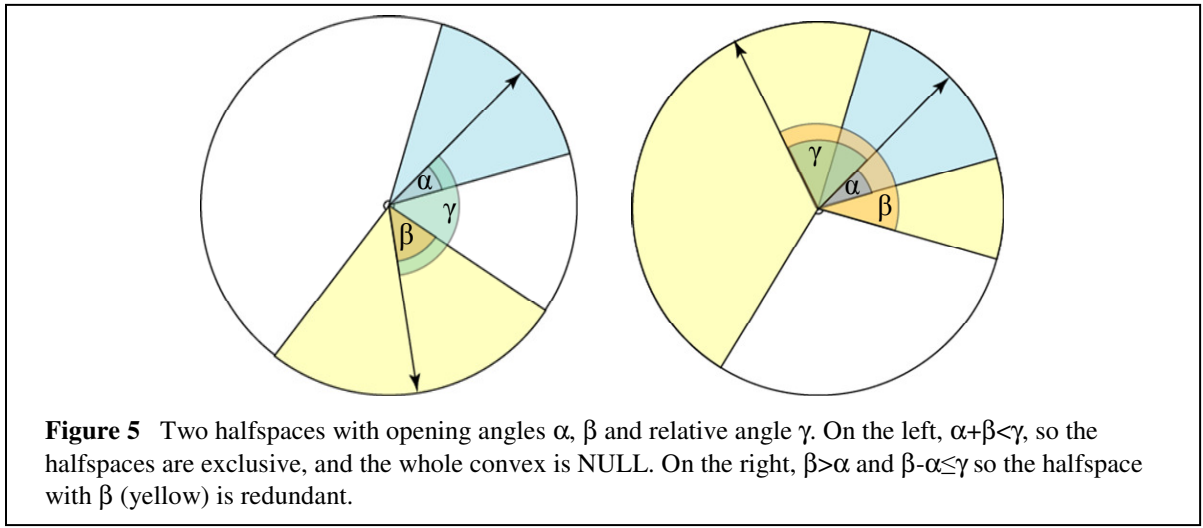

**Figure 5** Two halfspaces with opening angles $\alpha$, $\beta$ and relative angle $\gamma$. On the left, $\alpha+\beta<\gamma$, so the halfspaces are exclusive, and the whole convex is NULL. On the right, $\beta>\alpha$ and $\beta-\alpha\leq\gamma$ so the halfspace with $\beta$ (yellow) is redundant.

### Recognizing the NULL Convex

It is necessary to quickly decide whether the convex is empty, and if some halfspaces are redundant and can be eliminated. Consider a pair of halfspaces $h_1=\{n_1,d_1\}$ and $h_2 = \{n_2,d_2\}$, with arcangles of $\alpha = \mathrm{acos}(d_1)$, and $\beta = \mathrm{acos}(d_2)$. The relative angle of the two normal vectors can be computed from their dot product as $\gamma = \mathrm{acos}(n_1 \cdot n_2)$. The trivial test whether the two halfspaces are mutually exclusive is the criterion (left side of Figure 5).

$$\gamma \geq \alpha + \beta. \qquad (3.7)$$

In this case, we do not have to look further. The entire convex can be rejected as NULL.

*Eliminating Redundant Halfspaces*

A similar argument can be applied to the case where one halfspace can be dropped because one halfspace is contained inside another. If we assume that $\beta \geq \alpha$, then the figure on the right in Figure 5 shows that if:

$$\beta - \alpha \geq \gamma. \tag{3.8}$$

then β *contains* α, and so the β halfspace can be dropped. If there are $N$ halfspaces in a convex, this process consists of computing the γ of each pair of halfspaces and comparing it to their $\beta\text{-}\alpha$. It is an $O(N^2)$ computation, but, because $N$ is typically 5 and generally less than 100, this is an inexpensive operation. This is especially true because the halfspaces are sorted by size, and we compare the smallest to the largest first, eliminating the most obvious cases first. Obviously, if the convex consists of a single halfspace, we perform the trivial rejection tests only (*iii, iv*).

## 3.5. Computing the Vertices (Roots) of a Convex

The next simplifications are based on computing the *roots*–the pair-wise intersections of each pair of halfspace planes with the unit sphere. If a convex has no roots, it is either a single circle (a cap) or a circle with a hole in it (a belt). Both of these cases are simplified by the logic of section 3.4.

Assuming that there are, in fact, some roots of the convex, each halfspace pair defines 0, 1, or 2 points on the sphere. A root is *visible* if it is inside all the convex's halfspaces. If there are $N$ halfspaces in the convex, this is an $O(N^2)$ algorithm. We assume $N$ is typically 5 and is bounded by 100, so the approach outlined here is feasible. Knowing the visible roots allows us to discard redundant halfspaces, to compute patches, and to compute other properties of the convex.

Consider two halfspaces of the convex, described by $(x_1,y_1,z_1,c_1)$ and $(x_2,y_2,z_2,c_2)$. If the line defined by the intersection of these two planes pierces the unit sphere, then those line-sphere intersection points are roots.

The intersection of the two planes is defined by simultaneous solution of the two plane equations:

$$\mathbf{n}_1 \cdot \mathbf{x} = c_1, \qquad \mathbf{n}_2 \cdot \mathbf{x} = c_2 \tag{3.9}$$

We seek the equation of the line in the following form:

$$\mathbf{x} = u\mathbf{n}_1 + v\mathbf{n}_2 + w(\mathbf{n}_1 \times \mathbf{n}_2) \tag{3.10}$$

Here, the first two terms give the shortest vector from the origin to the line. The cross product $\mathbf{n}_1 \times \mathbf{n}_2$ is a vector normal to both $\mathbf{n_1}$ and $\mathbf{n_2}$, and is therefore a vector in the normal plane of both (see Figure 6). Its length is $\sin\theta$, where $\theta$ is the angle between the two vectors. So it is the vector parallel to the intersection line of the two planes. If we multiply Eq. (3.10) with $\mathbf{n}_1$ and $\mathbf{n}_2$, respectively, we get two scalar equations. Using the fact that (i) the two $\mathbf{n}$ vectors are unit vectors ($\mathbf{n_1 \cdot n_1} = \mathbf{n_2 \cdot n_2} = 1$), (ii) that they are normal to $\mathbf{n_1 \times n_2}$ so that their product with it is zero, and (iii) defining $\mathbf{n_1 \cdot n_2} = \cos\theta = \gamma$, we get:

$$u + v\gamma = c_1; \qquad u\gamma + v = c_2 \tag{3.11}$$

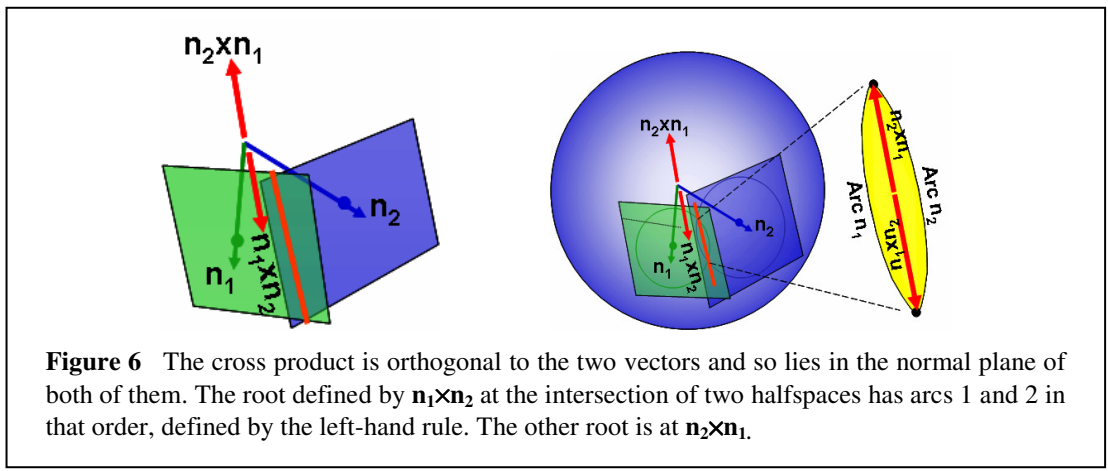

**Figure 6** The cross product is orthogonal to the two vectors and so lies in the normal plane of both of them. The root defined by $\mathbf{n}_1 \times \mathbf{n}_2$ at the intersection of two halfspaces has arcs 1 and 2 in that order, defined by the left-hand rule. The other root is at $\mathbf{n}_2 \times \mathbf{n}_1$.

One necessary condition for the solution is that the two planes should not be parallel, i.e., $\gamma \neq \pm 1$. We have to explicitly test for this condition, and exclude the pair of halfspaces from further consideration in the roots. Solving for $u$ and $v$, we obtain

$$u = \frac{c_1 - c_2\gamma}{1 - \gamma^2}; \qquad v = \frac{c_2 - c_1\gamma}{1 - \gamma^2} \tag{3.12}$$

Next, we need to find the points $\mathbf{x}$, which are on the unit sphere, given by $\mathbf{x \cdot x} = 1$. When we square the vector $\mathbf{x}$, we can again use the fact that $\mathbf{n}_1$ and $\mathbf{n}_2$ are unit vectors, and the fact that $|\mathbf{n_1 \times n_2}|^2 = \sin^2\theta = 1 - \cos^2\theta = 1 - \gamma^2$. The result of squaring Eq.(3.10) is

$$\mathbf{x \cdot x} = u^2 + v^2 + 2uv\gamma + w^2(1 - \gamma^2) = 1 \tag{3.13}$$

Now we need to solve this for $w$.

$$w^2 = \frac{1 - (u^2 + v^2 + 2uv\gamma)}{1 - \gamma^2} = \left(1 - \frac{c_1^2 + c_2^2 - 2c_1c_2\gamma}{(1 - \gamma^2)}\right)\frac{1}{1 - \gamma^2} \tag{3.14}$$

It is enough to consider only the positive root, because if we interchange the two normals, their vector product changes sign. So we can get the two different roots by intersecting plane A with B, and the other by intersecting B with A. The necessary conditions to have valid roots are

$$(1 - \gamma^2) > 0; \qquad (1 - \gamma^2) \geq c_1^2 + c_2^2 - 2c_1c_2\gamma, \tag{3.15}$$

where the equality means that we have a single root; not likely in floating point arithmetic.

This gives a maximum of *N(N-1)* possible roots. Now we need to determin, which of these roots are *good*; i.e., they satisfy all the other *(N-2)* halfspaces. We mark each root that does not satisfy one of the halfspace halfspaces as *bad*. The unmarked (good) roots are the vertices of the convex.

## 3.6. *Necessary Halfspaces and Masking Circles*

To compute the minimal set of halfspaces needed to define the convex, first compute the set, *S*, of all halfspaces that contribute to a good root (each good root contributes two halfspaces). The halfspaces in S are necessary to define the convex boundaries. But this set, *S*, of good circles is not a complete list. In addition to defining the good roots, it might also create a set of roots that must be excluded, because they do not appear on the list of good roots. In this case, some of the bad roots might have been masked by halfspaces not yet in the good circles list. We need to identify which halfspaces should be added to *S* to mask the bad roots.

The good circles set, *S,* is extended as follows. Make a list with an entry for each root and a flag called `good`. Initially, good roots have `good=1`; see green dots in Figure 7. A root has `good=0` if it is formed by two halfspaces from the good circles, but it is masked by another good circle. Roots, which are not formed by two good circles, have `good=-1`. The last possibility is to have a root, which is formed by two good circles, but it is masked only by a halfspace outside *S*. We mark these roots temporarily with `good=-2`. They are marked in blue in Figure 7. In the end, there will be no roots with `good=-2`.

We iterate though the roots with `good=-2`, and find the halfspace that eliminates most of them. Then we add that halfspace to good circles. We change the status of the roots involved to `good=0`. We iterate until there are no more roots left with a flag of –2. The closed set means that all intersections of the circles on the list are correctly masked. We call these halfspaces added later as the masking circles. For example, we would add the masking circle (red) in Figure 7. during this process.

All other halfspaces that are related to roots, beyond the ones on this list, are redundant and can be removed. We are left with non-intersecting holes. Of those, we need to determine which ones are visible, keep those, and delete the others.

### *Removing Additional Halfspaces*

Loop through the circles that do not have good roots. At this point, they still have an unknown state. These circles must be entirely inside or entirely outside of the convex because they have no good roots.

We can remove all holes (sign() < 0) fully outside the convex. The easiest test for this is to see if the hole's anti-center is outside the convex. Because nested holes have been eliminated, we keep inside holes and remove the rest.

### *Determining if a Halfspace Is Visible*

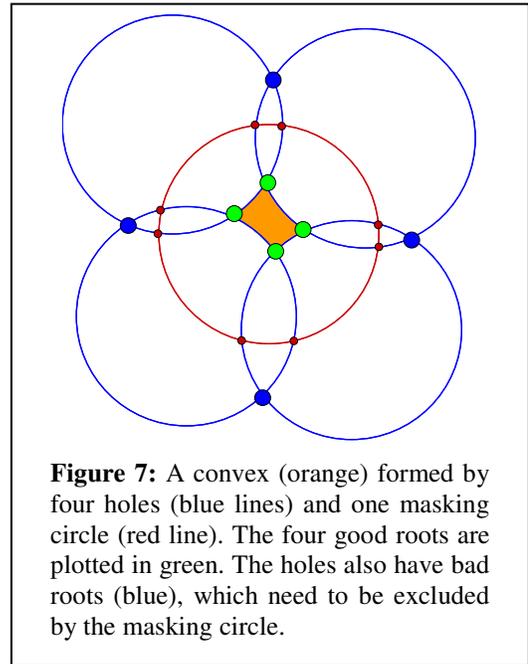

The *visibility test* arises when a halfspace is not involved in any of the roots. The halfspace can be either a bounding circle or a hole. It is either inside or outside the convex in its entirety. To test which of these two cases is valid, we need to pick an arbitrary point on its perimeter and test it against the convex.

We will pick a point west of the center of the circle. If the angular coordinates of the center are given by ($\alpha,\delta$), then we can write the normal vector **n** and the westward normal vector **w** in terms of these as

$$n_x = \cos\delta\cos\alpha \qquad w_x = -\sin\alpha$$
$$n_y = \cos\delta\sin\alpha \qquad w_y = \cos\alpha$$
$$n_z = \sin\delta \qquad w_z = 0$$
(3.16)

**Figure 7:** A convex (orange) formed by four holes (blue lines) and one masking circle (red line). The four good roots are plotted in green. The holes also have bad roots (blue), which need to be excluded by the masking circle.

It is clear that the two vectors are orthogonal. If the opening angle of the halfspace is $\cos\theta = \gamma$, then the vector of the point exactly west of the center is given by

$$\mathbf{r}_w = \mathbf{n}\cos\theta - \mathbf{w}\sin\theta. \qquad (3.17)$$

These points are pre-computed for each halfspace; since they are used in several contexts later (see patches). The visibility test for the nonintersecting circles consists of testing whether this point is contained in the convex. For these circles, if any point is inside, then all points are inside the convex.

### *Visible Halfspaces*

For each bounding circle, we do the visibility test and mark the result, but we keep the halfspace. For holes, we determine the visibility by using the test above and remove holes that are not visible.

## 3.7. Determining Arcs and Patches

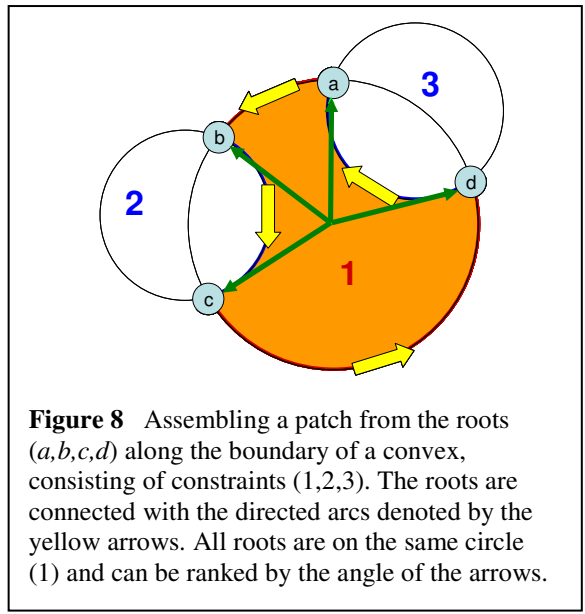

**Figure 8** Assembling a patch from the roots (*a,b,c,d*) along the boundary of a convex, consisting of constraints (1,2,3). The roots are connected with the directed arcs denoted by the yellow arrows. All roots are on the same circle (1) and can be ranked by the angle of the arrows.

Recall that a patch is a contiguous area of a convex – the convex of Figure 4 has eight triangular patches corresponding to the eight corners of the cube. What is the geometry of patches? We have a table of visible roots, which contains the position of the root, and the halfspaceID of the two parent circles defining the root. It is easy to invert this table into a table of arcs, which contain the halfspaceID, and the rootID of the beginning and the end of the arc, as long as the halfspaceID appears only twice in the root list. But, the visible roots table can be very degenerate. One halfspace can have many different pairs of roots (halfspace 1 in Figure 8 has four roots). In this case, additional computations, described later in this section, decide which pairs of roots define arcs.

Figure 8 shows the roots of a convex connected by directed arcs, denoted with the yellow arrows. The Roots table has four entries shown in Table 1. The first column of the table identifies a root. The second column contains the halfspaceID of the arc leading into the point. The third column has the halfspaceID of the arc leaving the vertex.

From Table 1, it is obvious that there is a single arc for halfspace 2, starting with *b* and ending in *c*, written as [2:*b,c*], and that there is another one for halfspace 3, starting with *d* and ending in *a*, represented as [3:*d,a*]. So, we can copy these two rows into the arcs table. In general, if a halfspace *A* has one root *r*, then add [*A: r, r*] to Arcs. If it has two roots, *r1, r2* then add [*A: r1, r2*] or [*A: r2, r1*] to Arcs. However, we see that all four of the roots are related to halfspace 1, and it is not clear which root is connected to which.

To solve this degeneracy and order (*r1, r2*), we impose a counterclockwise ordering of the roots and arcs within the same circle. This is done by transforming the problem into the plane of the halfspace, as shown in Figure 6. Calculating the angle (measured from north) of the green position vectors connecting the center of circle C1 to each of the four roots, the counterclockwise rank ordering of the roots would be (*b,c,d,a*). Because *c* is the beginning of an arc of C1, and the consecutive *d* is an end, it is clear that [1:*c,d*] is an arc. The other connected arcs are also easily obtained. The resulting arcs table is shown in Table 2.

**Table 1** Roots

| | | |
|---|---|---|
| b | 1 | 2 |
| a | 3 | 1 |
| d | 1 | 3 |
| c | 2 | 1 |

**Table 2** Arcs

| | | |
|---|---|---|
| 2 | b | c |
| 3 | d | a |
| 1 | c | d |
| 1 | a | b |

How can we compute the projection of the roots on the plane of the circle? Each root can be expanded as the linear combination of three normal vectors, the normal vector **n**, the westward normal **w**, and the northward vector **u**. This latter vector can be obtained as $\mathbf{u} = \mathbf{w} \times \mathbf{n}$. When this is done, we can write the root **r** as

$$\mathbf{r} = \mathbf{n}\cos\theta + (\mathbf{u}\cos\phi + \mathbf{w}\sin\phi)\sin\theta. \qquad (3.18)$$

As a result, we can get the lateral angle $\phi$ from the arctan of the **u**, **w** coefficients.

$$\cos\phi\sin\theta = \mathbf{u}\times(\mathbf{r}-\mathbf{n}\cos\theta) = \mathbf{u}\times\mathbf{r},$$
$$\sin\phi\sin\theta = \mathbf{w}\times(\mathbf{r}-\mathbf{n}\cos\theta) = \mathbf{w}\times\mathbf{r},$$
(3.19)

because $\mathbf{u}$ is perpendicular to $\mathbf{n}$. By increasing $\phi$ to sort the roots in a circle, we get the required counterclockwise ordering, and we know which segments are visible because the start and end roots are known. These visible arcs are then added to the *arcs* list.

When we have iterated over the list of all circles, we have a unique list of all arcs. These are still not ordered in the correct fashion: they do not form closed patches.

*Assembling Arcs into Patches*

The next step is to assemble the arcs into closed patches. We have to go in a particular order because we want to be able to use the same output list of arcs to drive the visualization. The inside of a visualized polygon depends on the winding number; therefore, we first have to output the bounding circles, followed by the rooted halfspaces, ending with the unrooted holes.

**Table 3**  Patch

| 1 | a | b |
| 2 | b | c |
| 1 | c | d |
| 3 | d | a |

*Rooted Halfspaces*

We start with the smallest rootid and get the arc that starts with this root. We then enter a loop, where the next root is picked from the endpoint of the previous arc. When we reach the end, i.e., we get back to the root we started with, we increment the patch number. Again, get the smallest rootid from the remainder list and loop. Following our example from Figure 8, we would pick the arc that starts with *a*. Because this arc ends in *b*, the next one will have to start in *b*. Because that ends in *c*, the next one starts with *c*, and so on. The resulting patch is seen in Table 3.

*Unrooted Holes*

At the end, we output the circles making up the enclosed holes. They are written in the same form as the bounding circles; the westward point is the beginning and end point of the arc.

## 3.8. Simplification Using the Arcs

Given the ordered list of arcs, including the masking circles and holes, we can create a list of the halfspaces (parents of these arcs) that contribute to the convex. Other halfspaces are redundant and can be removed. This final step completes the simplification process. The list of arcs is also kept for subsequent visualization purposes.

# 4. Spatial Queries: Intersecting with the HTM

Computing a list of trixels that cover a region is the most complex aspect of the HTM implementation. We want to compute the HTM trixels that are either inside or partially inside a given region. We will refer to *full* and *partial* trixels, depending whether the trixel is fully or partially contained in the region.

## 4.1. Intersecting a Trixel with a Halfspace

First, we look at the intersection of a trixel $t=\{\mathbf{v}_i\}$ with a halfspace $c=\{\mathbf{v}_c;d\}$. Figure 9 shows the decision tree of intersecting *c* with a single HTM trixel. The first test is whether the three corners $\mathbf{v}_i$ of the trixel node are inside or outside *c*. If

$$\mathbf{v}_c \cdot \mathbf{v}_i > d \qquad (4.1)$$

then the corner *i* of the trixel is inside the *c* halfspace. If the halfspace is positive, and all three corners are inside, the whole trixel must be inside the halfspace (mark the trixel as full). This uses the fact that the trixels are bounded by great circles. If some corners are inside and some outside, the halfspace's boundary has to pass through the node, and we get a partial trixel. If all corners are outside, we need to test further because the trixel might intersect with the halfspace at other places. We certainly will not have a full trixel at that point any more, but we want to be sure whether *c* is outside or partially covering the trixel.

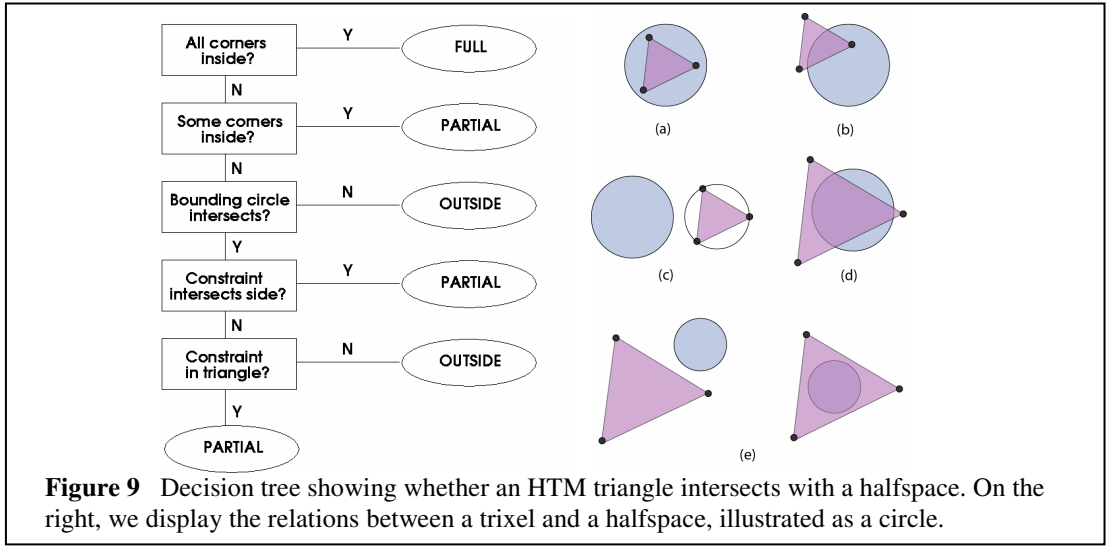

**Figure 9** Decision tree showing whether an HTM triangle intersects with a halfspace. On the right, we display the relations between a trixel and a halfspace, illustrated as a circle.

For the next test, we calculate a bounding circle of the trixel, which is given by the intersection of the sphere with the plane of the triangle. As always, we call the three corners of the triangle ($\mathbf{v}_0, \mathbf{v}_1, \mathbf{v}_2$); ordered counterclockwise. *b* is defined by

$$\begin{aligned} b &:= \{\mathbf{v}_b; d_b\} \\ \mathbf{v}_b &= \frac{(\mathbf{v}_1 - \mathbf{v}_0) \times (\mathbf{v}_2 - \mathbf{v}_1)}{|(\mathbf{v}_1 - \mathbf{v}_0) \times (\mathbf{v}_2 - \mathbf{v}_1)|} \\ d_b &= \mathbf{v}_0 \cdot \mathbf{v}_b \end{aligned} \qquad (4.2)$$

We test whether the halfspaces *b* and *c* overlap by comparing the sum of the angles $\varphi_c + \varphi_b$ to the angle $\theta$ between $\mathbf{v}_c$ and $\mathbf{v}_b$. If $\theta$ is less than the sum, we have an overlap. If the halfspaces *c* and *b* do not intersect, the triangle lies outside the halfspace (the third decision box in Figure 9).

If *c* and *b* do intersect, we test whether the halfspace actually intersects with one of the triangle sides. If yes, we mark the triangle as "partial." If not, we still have to determine whether the halfspace lies fully inside the triangle or fully outside. We do this by checking whether $\mathbf{v}_c$ is inside $\mathbf{v}_0, \mathbf{v}_1, \mathbf{v}_2$. If

$$(\mathbf{v}_i \times \mathbf{v}_j) \cdot \mathbf{v}_c < 0 \qquad (i,j) \in \{(0,1);(1,2);(2,0)\} \qquad (4.3)$$

for any allowed pair *(i,j)* then *c* is outside (at this point we make use of the fact that the three corners are ordered counterclockwise). If all three tests in Eq.(4.3) are false, *c* lies completely inside the triangle and we mark it "partial." This concludes the decision tree of Figure 9.

## 4.2. The Intersection of a Halfspace with an Edge of a Triangle

Any edge of a triangle is given by its end points $\mathbf{v}_1, \mathbf{v}_2$. They are connected by a great circle segment. Because this is not unique, we specify that they are always connected by the shorter of the two possible great circle segment connections. For HTM nodes, this is always true because depth 1 (the largest trixel) has sides that are ¼ of a great circle. All three corners of a trixel always lie in the same half-sphere. To see whether a halfspace intersects an edge, we first parameterize the great circle segment connecting $\mathbf{v}_1$, $\mathbf{v}_2$:

$$\vec{e}(\vartheta) = \mathbf{v}_1 \cdot \frac{\sin(\theta - \vartheta)}{\sin(\theta)} + \mathbf{v}_2 \cdot \frac{\sin(\vartheta)}{\sin(\theta)}, \qquad (4.4)$$

where $\vartheta$ runs from 0 to $\theta$, the angle between $\mathbf{v}_1$ and $\mathbf{v}_2$. This equation can be rewritten as

$$\vec{e}(s) \cdot (1 + s^2 u^2) = \mathbf{v}_1 (1-s)(1+u^2 s) + \mathbf{v}_2 s(1+u^2), \qquad (4.5)$$

with

$$u = \tan(\theta/2) \quad ; \quad s \cdot u = \tan(\vartheta/2), \qquad (4.6)$$

i.e. $s$ runs from 0 to 1. Now, we only need to test whether any of the points $\mathbf{e}(s)$ lie in the plane that defines $c$. We get the following equation for $s$:

$$\mathbf{v}_c \cdot \mathbf{e}(s) - d = 0 \qquad (4.7)$$

Substituting Eq.(4.5) in the previous equation, we get a quadratic equation in $s$:

$$-u^2(\gamma_1 + d) s^2 + (\gamma_1(u^2 - 1) + \gamma_2(u^2 + 1)) s + \gamma_1 - d = 0 \qquad (4.8)$$

We wrote $\gamma_i = \mathbf{v}_c \cdot \mathbf{v}_i$ for the two edges $i=1,2$. Solving this equation for $s$ tells us how many times the triangle edge $\mathbf{v}_1, \mathbf{v}_2$ intersects the halfspace cap $c$, and how many solutions are in the range (0,1).

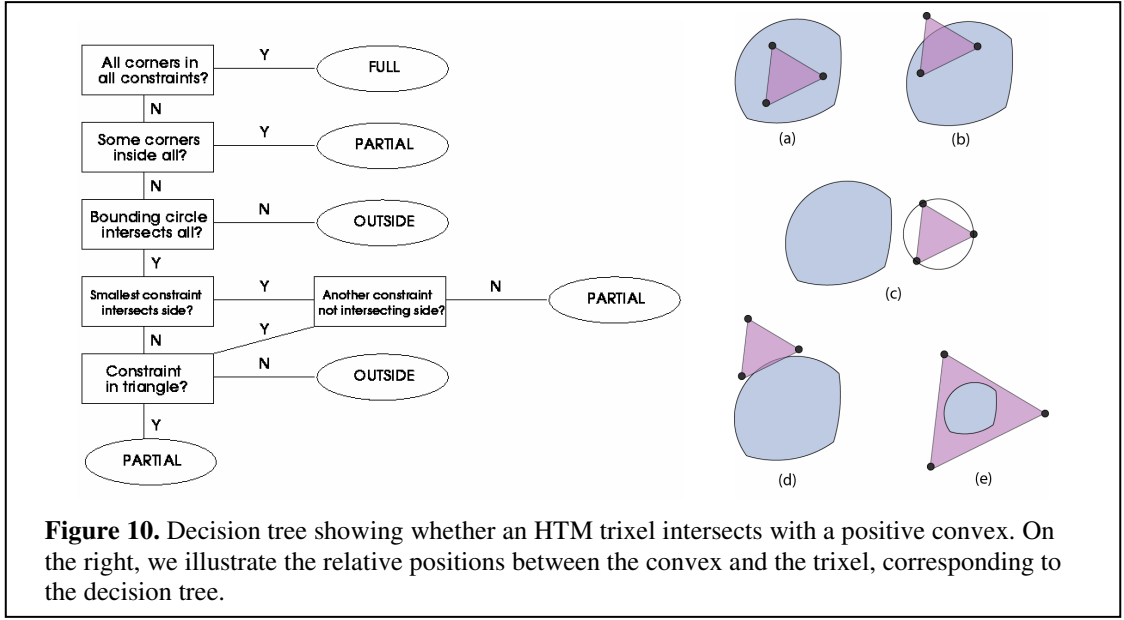

**Figure 10.** Decision tree showing whether an HTM trixel intersects with a positive convex. On the right, we illustrate the relative positions between the convex and the trixel, corresponding to the decision tree.

## *4.3. Intersecting a Trixel with a Convex*

Intersecting a trixel with a convex (the intersection of several halfspaces) is more complicated. Depending on the sign of the convex (Eq. 3.5) different procedures determine whether a triangle lies fully or partially inside.

### *i) Positive and Zero Convexes*

If the sign of the convex is +1 or zero, the area it defines is a convex patch on the sphere. The decision tree for a trixel is almost identical to the decision tree for a simple halfspace (Figure 10). The only complication is that we need to test each halfspace for intersection with the trixel to be sure whether it is partial or outside.

### *ii) Negative Convexes*

The decision tree of the positive convexes is not applicable for negative convexes at all. We have another decision tree, displayed in Figure 11. If all three vertices are inside all halfspaces of the negative convex, we still need to test whether there is a hole inside the triangle. We do this by testing each halfspace's vector $\mathbf{v}_c$ in the convex to determine whether it is inside the node. If there is such a halfspace, we can flag the triangle as partial. If there is none, test whether one of the negative halfspace's boundary circles intersects with one of the edges of the triangle. If yes, flag the triangle as partial. If not, it is full. If just one or two corners are inside all halfspaces, then we can safely assume the triangle as partial. This step is the same for all convexes.

If all corners are outside the convex, we still need to investigate further whether the triangle is partial or outside: If none of the halfspaces intersect with the triangle edges, the triangle can be safely assumed to be outside the convex. If there are intersections, we might have very complicated patterns inside the triangle (Figure 11, bottom), so we just assume partial to be certain. Complicated calculations can be applied to sort all special cases out, but the computational cost is usually not worth the effort.

### *iii) Mixed Convexes*

Mixed convexes are those with both positive and negative halfspaces. We can imagine a positive convex area on the sphere where we chip off some of it (there might also be holes). The decision tree used for this case is a mix of the positive and negative decision trees.

There has to be at least one positive halfspace, and if the triangle lies fully outside any positive halfspace in the mixed convex, it is certainly outside the whole convex. Also, if only one or two of the three corners are inside the convex (all halfspaces) and the other corners are outside, the triangle can be flagged as partial. If all three corners are inside all halfspaces and none of the halfspaces intersects the triangle sides, the triangle can safely be assumed fully inside the convex.

If none of the most common cases described apply, we usually flag the triangle as partial. We can get very rare and very complicated setups where we might be wrong, but again, the computational effort to determine whether the triangle is outside is usually not worth it.

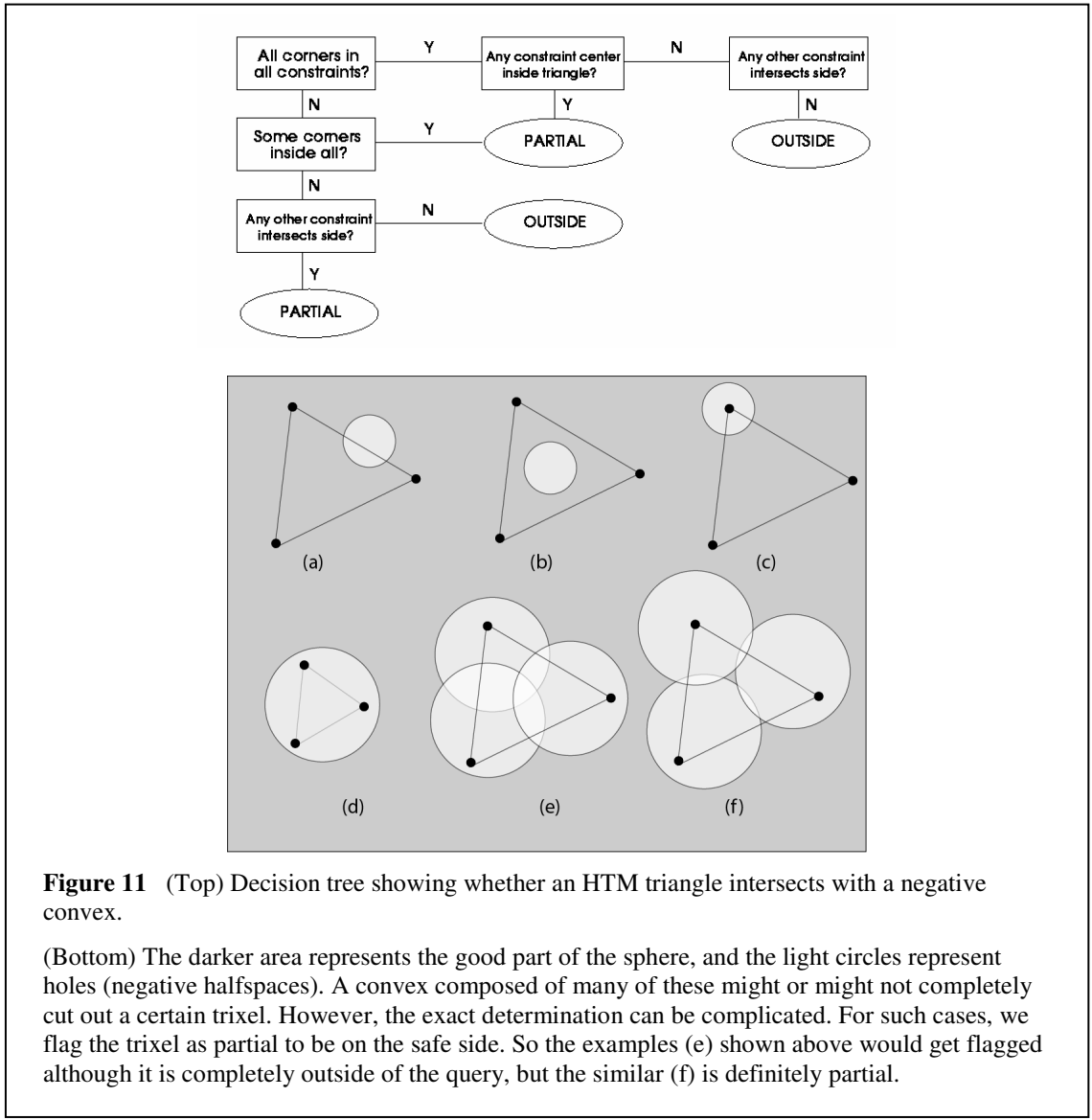

**Figure 11** (Top) Decision tree showing whether an HTM triangle intersects with a negative convex.

(Bottom) The darker area represents the good part of the sphere, and the light circles represent holes (negative halfspaces). A convex composed of many of these might or might not completely cut out a certain trixel. However, the exact determination can be complicated. For such cases, we flag the trixel as partial to be on the safe side. So the examples (e) shown above would get flagged although it is completely outside of the query, but the similar (f) is definitely partial.

## *4.4. Terminating the Recursion*

The intersection algorithm generates a list of trixels that are classified either full or partial and provides a reasonably tight cover around some given convex. Full trixels are always accepted into the list, because they are completely contained within the convex. But a partial trixel can be split into four subtrixels, and the same test is applied to each offspring. Thus, a recursive algorithm builds a list of trixels so that the fit gets tighter as the depth of recursion depth increases. In the real, less-than-ideal world, convex boundaries do not line up perfectly with the grid-lines of the HTM. Therefore, we can expect to have partial trixels at any depth. There needs to be a criterion that halts further recursion. One way is to set an absolute maximum depth. This guarantees the tightness of a fit to within the feature size of the trixels at this depth. The downside is that the number of trixels generated can still be impractically large. The following heuristic seems to work well in providing a balance between precision (goodness of fit) and computational effort.

The crux of the heuristic is the decision of whether to split a partial trixel before the maximum depth has been reached. It does that by looking into the trixel's four children and making a decision based on their classification and that of the parent. The recursion will always stop when it reaches the preset (tunable) maximum depth.

*Heuristic test*

Let $F$ and $P$ be the number of full and partial subtrixels, respectively. Let $P'$ be 1 + the number of partial siblings (partial subtrixels of the parent). We stop the recursion if any of these conditions are met:

| | |
|---|---|
| `(P==4)` | All children are partial |
| `(F>2)` | More than two children are full |
| `(P==3 && F==1)` | Three children are partial, and one is full |
| `(P>1 && P'==3)` | Parent had three partials, and more than one child is partial |

In our first implementation, our experience was that the recursion often stopped too early because the last condition in the above heuristic was met too often. The result was a cover of trixels that overshot the convex by a great deal. The way to remedy that was to ignore the last condition until the recursion reached a minimum depth, so that the trixel size was comparable to or smaller than the size of the convex.

## 4.5. *Translating Trixels to HTM Ranges*

The trixel cover consists of trixels at various levels. To be effectively usable for a database application, all trixels considered are on a given fixed level. Any trixel is equal to the sum of its children, each of which is equal to the sum of its four children, etc. So, any trixel is equivalent to a finite, albeit a potentially large, number of smaller trixels at a greater level. In our case, we chose depth 21 for practical considerations. The numbers formed by the bit-patterns of the trixel addresses that comprise a single ancestor are contiguous without gaps. Therefore, a trixel at, for example, depth *n* can be represented as a range of numbers, all of which represent some trixels at depth *(n+k)*. A complicated region that contains trixels at many different levels can be expressed as a set of (low, high) intervals.

Still, the potential exists for creating too many ranges. Some ranges produced are consecutive intervals, that is, *(a,b)* and *(b+1, c)*, which can be merged into a single interval *(a,c)*. If the number of ranges is still too high, then intervals that have the smallest gap are merged, such as when *k>1*, *(a,b)* and *(b+k,c)* are replaced by *(a,c)*. The price to pay for this is the implicit introduction of trixels represented by *{b+1, b+2,...b+k-1}*, which are not part of the computed cover, but are now part of an overshoot. Our algorithm allows the specification of the maximum number of intervals desired in the cover. The price of having overshoots is to have to do more fine-grain analysis in spatial searches, but it is preferred over undershoots. In this latter case, the region under consideration would not be covered by the trixels in the cover. As a result, those areas would not be considered during the spatial search. Because that would not be acceptable behavior, any overshoot is preferable to even the slightest undershoot.

# 5. Bounding Circles and Convex Hulls

As part of the spatial data package, we implemented several other algorithms that complement the basic HTM functions. These additions include the region algebra [12], the creation of convex hulls, and the computation of arc lengths, areas and optimal bounding circles. We describe the underlying geometric algorithms here. They are implemented by using a combination of SQL and C#.

*Computing the Bounding Circles*

We compute bounding circles for each patch, possibly more than one per convex, that can be used for fast intersection tests. To detect whether two convexes overlap, we just have to perform the intersection for those pairs of convexes where any of their bounding circles intersect, using Eq. (3.7).

Consider a single patch, with its ordered list of vertices. Only the vertices along drawable arcs are considered. A bounding circle will contain at least two, but possibly more, of its vertices, similar to the classic planar case. There is one major difference between the spherical polygons we are considering and the planar case: the arcs of our polygons can also be small circles, with either positive or negative curvature (where zero curvature is assigned to great circles, the geodesic curves on the surface of the sphere). Concerning the bounding circle, we can ignore the negative small circles, and just assume that the two endpoints are connected with a great circle. If all points of the great circle are contained within the bounding circle, then all points of the small circle arc will also be inside.

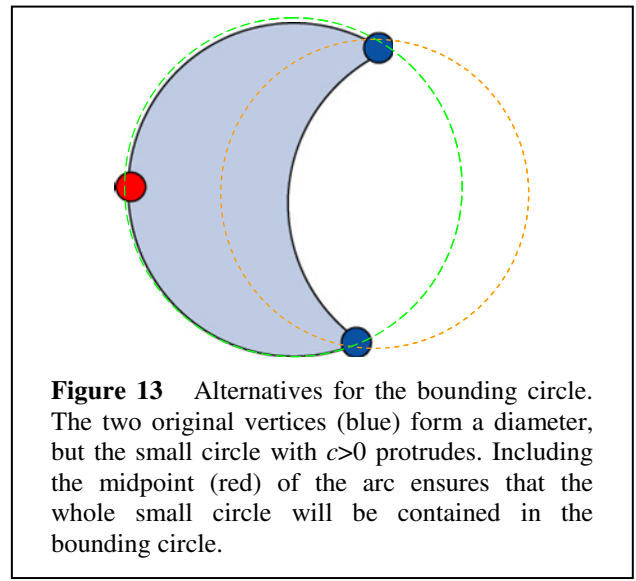

It is trickier when we have a positive small circle arc that protrudes, as compared to a great circle. In this case, we will introduce an additional vertex point at the midpoint of the arc (Fig.13). We force the midpoint of the arc to be in the plane of the small circle half-space halfspace.

First, we save the smallest circle in the convex as candidate (i) for the bounding circle. This will be compared to other alternatives. If there are exactly two vertex points on the bounding circle, then they must form a diameter, similar to the planar case. We test all pairs of the original vertices and pick the largest distance. Then, we test all other (extended) vertex points. If all the vertices are inside, we accept the solution as a possible bounding circle (ii), where the center is at the midpoint of the great circle connecting the two endpoints, and its diameter is determined from the distance of the points.

**Figure 13** Alternatives for the bounding circle. The two original vertices (blue) form a diameter, but the small circle with $c>0$ protrudes. Including the midpoint (red) of the arc ensures that the whole small circle will be contained in the bounding circle.

Next, we try all possible triplets formed out of the extended vertices. We compute the equation of the plane going through each, using the winding order of the vertices to define the direction of the normal. We pick the circle with the smallest radius (iii). Next, we select the smallest circle of the set of (i), (ii), and (iii) as the final choice for the bounding circle.

*Computing the Area*

It is often necessary to know the area of a spherical polygon. Girard's formula gives a closed expression for a "real" spherical polygon, with all edges formed by great circles. Our case is more complicated, because many of the edges of our polygons will be small circles. We employ a trick also advocated by Song, et al [11]. The idea is to take the vertex points of our generalized spherical polygon and connect them with great circles. Our polygon and the new one defined by the great circles differ only in small lenticular areas at around the perimeter. These areas, referred to as semilunes, are bounded by one great circle and one small circle. Their area is signed, depending on the sign of the small circle. If we can calculate the areas of the semilunes, we have almost solved the problem.

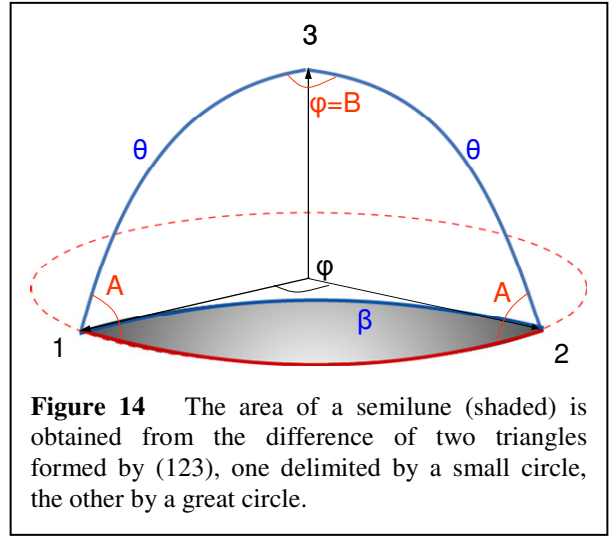

**Figure 14** The area of a semilune (shaded) is obtained from the difference of two triangles formed by (123), one delimited by a small circle, the other by a great circle.

To compute the area of the semilunes, let us rotate our coordinate system to one where the normal to the plane of the small circle is our *z*-axis (Fig. 14). Let us add the +*z* pole of the coordinate system as a new vertex to the other two. Now we have two triangles, one fully bounded by great circles, the other bounded by two great circles and one small circle. The original vertex points have spherical coordinates $(\theta,\varphi_1)$ and $(\theta,\varphi_2)$ in the new spherical coordinate system, where $\cos\theta = c$, the offset of the small circle plane from the origin. The area of the whole cap sliced off by the small circle is $2\pi(1-\cos\theta)$. The area of the small circle triangle is prorated by the fraction of $2\pi$ covered by $\varphi=\varphi_2-\varphi_1$:

$$A_s = \varphi(1-\cos\theta). \tag{5.3}$$

The computation of the angle $\varphi$ is described in Eq.(5.12).

The area of the great circle triangle can be computed from the three angles at the vertices, using Girard's formula of spherical excess as

$$A_g = 2A + B - \pi, \tag{5.4}$$

because the vertex angle at the pole is $B=\varphi$ and the other two angles A are the same, dependent only on $(\theta,\varphi)$. The great circle distance of points (1,3) and (2,3) is $\theta$, and of points (1,2) is

$$\beta = 2\operatorname{asin}\frac{|\vec{x}_2 - \vec{x}_1|}{2}. \tag{5.5}$$

After some spherical trigonometry, *A* can be expressed as

$$\cos A = \frac{\tan\beta/2}{\tan\alpha}, \tag{5.6}$$

The area of the semilune is the difference of the two triangles:

$$A_{sl} = A_s - A_g = 2(\frac{\pi}{2} - A) - B\cos\alpha. \tag{5.7}$$

We have deliberately used ($\pi/2-\alpha$) because the inverse trigonometric function *asin* is much more accurate and stable than *acos* for very small angles, thus

$$\frac{\pi}{2} - A = \operatorname{asin}\left(\frac{\tan\beta/2}{\tan\alpha}\right). \tag{5.8}$$

To compute the full area of the polygon, consider the center of the bounding circle O, and the original vertices labeled 1..N. Let us denote the area of the great circle triangle formed by vertices ($O,n,n+1$) as $\Delta_n$, which can be computed from Girard's formula, and the area of the semilune corresponding to the arc ($n,n+1$). The area of our polygon describing the patch is

$$A_{poly} = \sum_{n=1}^{N}(\Delta_n + S_n) \qquad (5.9)$$

The area of the whole region is obtained through another sum over all patches that belong to the region. The whole algorithm is implemented through three functions.

*Computing the Arc Length*

Sometimes we will need to proper length the arcs forming the edges of the polygons. These will be different from the great circle distances. This is calculated via transforming into the system of the small circle, and evaluating Eq.(5.2), and then taking the difference of the azimuthal angles in the plane of the small circle.

*Convex Hulls*

Convex hulls for a set of points on the surface of a sphere are only well defined if the points all fit within half of the globe. This is a zero-halfspace, or a halfspace, for which d=0. Our algorithm for computing the convex hull is based on projecting the points from the hemisphere that contain them onto a plane tangent to the halfspace at its center. The problem is then reduced to 2D, where any convex hull algorithm works. We are using one due to Andrews [15].

# 6. Implementation and Performance

These HTM algorithms were originally implemented in C, C++, and Java. This code has been widely used by the astronomy community. In 2000, we adapted the C++ library to work with SQL Server 2000 as a set of external stored procedures. The various implementations can be downloaded from http://www.sdss.jhu.edu/htm. The C++ implementation has been used in the Sloan Digital Sky Survey [16], Virtual Sky [17], SkyQuery [18], SuperCOSMOS Sky Surveys [19], and the STScI Guide Star Catalog 2 [20].

|  | *SQL2005 + C++* | *SQL2005 + C#* |
|---|---|---|
| `fHtmLatLon()` | 1.143 ms | 0.019 ms |
| `FHTMCover()` | 4.100 ms | 0.710 ms |
| `fGetNearbyObjects()` | 8.800 ms | 1.574 ms |

**Table 4** The CPU cost (in milliseconds) of the various implementations of the HTM package. The regions used for fHtmcoverRegion() were complex, chosen from the Sloan Digital Sky Survey, and the fHtmNearbyObjects was computed for stream gauges within 30 nautical miles of each city. Most of the increased performance is due to the improved function linkage of common language runtime (CLR) table-valued functions. The dominant cost is now the cost of the HTM code itself.

The performance of the C++ implementation on a 1 GHz Intel processor is shown in Table 4. It was measured by evaluating the `fHtmEq()` function in both libraries, and by evaluating the `fHtmCoverRegion()` function on 24,013 regions (28,466 convexes and 109,505 halfspaces) that define interesting footprints of the Sloan Digital Sky Survey Data Release 3. The SQL Server 2000 performance was tolerable, and allowed quite flexible spatial data queries of astronomy datasets. To improve performance, we rewrote the HTM library in C# and used the CLR binding in SQL Server 2005. As Table 4 shows, this increased performance by a factor of 50 for scalars, and by a factor of 6 for the table-valued functions. Most of the time is now in the spatial library.

There is a tradeoff in building an HTM cover of a region. Large trixels can be used inside the region, but many small trixels approximate the area boundary more accurately. However, many small trixels imply many trixel-range lookups in the spatial index. On our test computers, using the SQL Server 2005 B-tree index, the trixel range lookup is about five times more expensive than a distance test (a dot product). Assuming a uniform distribution, it might be more efficient to subdivide a trixel four ways, hoping that one of the four sub-trixels will be discarded (outside the region), and so ¼ of the points will not be examined and have their distance computed. If three edge trixels survive (three overlap the region), then the trixel lookup cost goes from five units to 3×5=15 units—an increase of 10 units to examine three, instead of one, HTM range. But discarding a trixel saved ¼ of the distance comparisons, each costing one unit. Therefore, the discarded sub-trixel must have at least 10 objects to break even. This implies that the object density indicates what the finest trixel granularity should be. The new C# library has this logic built into it. In the future, we hope to make the library adaptive so that fine-grain datasets will have more covering trixels to more closely approximate the region.

This logic generalizes, if the object density is $D$, and if the trixel area is $A$, and if the cost ratio of B-tree probe to an inside the region test is $R$, then subdivision of trixel area $A$ makes sense if the extra probes are less expensive than the saved comparisons: *3R<DA/4* or *A >12R/D*. As mentioned previously, on this implementation $R$=5. For the SDSS data, the density $D$ is approximately 40 objects/arcmin$^2$, so depth 13 (about 1.7 arcmin$^2$) is a good granularity for SDSS. Conversely, the USGS data here has a density of .01 objects per arcmin$^2$, and so depth 9 is a more appropriate limit for that application. In general, if $D$ is the density in objects/arcmin$^2$, then *12–log$_4$(R/D)* is a good maximum depth for the HTM mesh.

## 7. Summary

The HTM is a very quick and powerful method to implement spherical partitioning. It is based on a spherical quadtree mapped onto a B-tree index in SQL Server. Combined with a halfspace-based description of generalized spherical polygons, we can recursively compute a trixel cover of the polygon over the sphere, which is then easily converted to a set of range queries in a database. We found the performance of the algorithms acceptable even in astronomical databases containing several hundred million objects.